\documentclass[twocolumn]{aastex63}

\submitjournal{PSJ}
\accepted{June 18th, 2021}
\turnoffedit

\shorttitle{Earth-Sun Transit System}
\shortauthors{Mayorga, Lustig-Yaeger, May, et al.}

\graphicspath{{./}{figures/}}

\newcommand{\jwst}{{JWST}}
\newcommand{\hst}{\emph{HST}}

\newcommand{\hho}{H$_2$O}
\newcommand{\nno}{N$_2$O}
\newcommand{\coo}{CO$_2$}
\newcommand{\oo}{O$_2$}
\newcommand{\ooo}{O$_3$}
\newcommand{\chhhh}{CH$_4$}
\newcommand{\tdur}[1]{t$_{#1}$}
\begin{document}

\title{Transmission Spectroscopy of the Earth-Sun System to Inform the Search for Extrasolar Life}

\correspondingauthor{L. C. Mayorga}
\email{laura.mayorga@jhuapl.edu}

\author[0000-0002-4321-4581]{L. C. Mayorga}
\affiliation{The Johns Hopkins University Applied Physics Laboratory, 11100 Johns Hopkins Rd, Laurel, MD, 20723, USA}
\author[0000-0002-0746-1980]{J. Lustig-Yaeger}
\affiliation{The Johns Hopkins University Applied Physics Laboratory, 11100 Johns Hopkins Rd, Laurel, MD, 20723, USA}
\affiliation{NASA NExSS Virtual Planetary Laboratory, Box 351580, University of Washington, Seattle, Washington 98195, USA}
\author[0000-0002-2739-1465]{E. M. May}
\affiliation{The Johns Hopkins University Applied Physics Laboratory, 11100 Johns Hopkins Rd, Laurel, MD, 20723, USA}
\author[0000-0001-7393-2368]{Kristin S. Sotzen}
\affiliation{The Johns Hopkins University Applied Physics Laboratory, 11100 Johns Hopkins Rd, Laurel, MD, 20723, USA}

\author[0000-0002-9032-8530]{Junellie Gonzalez-Quiles}
\affiliation{Department of Earth and Planetary Sciences, Johns Hopkins University, 3400 N. Charles Street, Baltimore, Maryland 21218, USA}

\author[0000-0003-4220-600X]{Brian M. Kilpatrick}
\affiliation{Space Telescope Science Institute, Baltimore, MD 21218, USA}

\author[0000-0002-0618-5128]{Emily C. Martin}
\affiliation{Department of Astronomy \& Astrophysics, University of California, Santa Cruz, 1156 High
St. Santa Cruz, CA 95064}

\author[0000-0001-8397-3315]{Kathleen Mandt}
\author[0000-0002-7352-7941]{K. B. Stevenson}
\author[0000-0003-1629-6478]{N. R. Izenberg}
\affiliation{The Johns Hopkins University Applied Physics Laboratory, 11100 Johns Hopkins Rd, Laurel, MD, 20723, USA}

\begin{abstract}
Upcoming NASA astrophysics missions such as the James Webb Space Telescope will search for signs of life on planets transiting nearby stars. Doing so will require co-adding dozens of transmission spectra to build up sufficient signal to noise while simultaneously accounting for challenging systematic effects such as surface/weather variability, atmospheric refraction, and stellar activity. To determine the magnitude and impacts of both stellar and planet variability on measured transmission spectra, we must assess the feasibility of stacking multiple transmission spectra of exo-Earths around their host stars. Using our own solar system, we can determine if current methodologies are sufficient to detect signs of life in Earth's atmosphere and measure the abundance of habitability indicators, such as \hho{} and \coo, and biosignature pairs, such as \oo{} and \chhhh. We assess the impact on transmission spectra of Earth transiting across the Sun from solar and planetary variability and identify remaining unknowns for understanding exoplanet transmission spectra. We conclude that a satellite  observing Earth transits across the Sun from beyond L2 is necessary to address these long-standing concerns about the reliability of co-adding planet spectra at UV, optical, and infrared wavelengths from multiple transits in the face of relatively large astrophysical systematics.
\end{abstract}


\section{Introduction} \label{sec:intro}
Over the next decade, users of the James Webb Space Telescope (\jwst) will apply transit-based techniques to determine whether rocky planets orbiting M-dwarf stars have tenuous, clear, or cloudy atmospheres. Through secondary eclipse measurements, \jwst{} users will then begin identifying which of those terrestrial planets that reside within the habitable zone \citep{kopparapu_et_al2013}, where planetary surface temperatures are suitable for liquid water, are actually habitable. Finally, and most difficult, astronomers will analyze transmission spectra measurements in the search for signs of life. \edit1{Studies, like \citet{Fauchez2019, Lustig-Yaeger2019, Wunderlich2019, Tremblay2020}, have shown that }\jwst{} will need to accumulate many dozens, if not hundreds, of transits over its 5.5-year primary mission to build up sufficient signal to potentially confirm the presence of biosignature pairs such as \oo{} and \chhhh, \ooo{} and \nno, and others \deleted{Wunderlich2019, Lustig-Yaeger2019, Fauchez2019} \edit1{\citep[see][and references therein]{Krissansen-Totton2016}}. Such an observation campaign will require a substantial investment in telescope time and, thus, will likely only be attempted for a small number of targets within the mission's lifetime. While the search for signs of extrasolar life may also become possible using other techniques, \deleted{such as  direct imaging, future large mission concepts such as the \emph{Habitable Exoplanet Observatory}  and the \emph{Large UV Optical Infrared telescope}  are likely two decades away and the next-generation of ground-based telescopes (using instruments such as METIS or PCS) have ambitious planet-star contrast and separation requirements for even the nearest M dwarfs, but will have optical wavelength coverage that \jwst{} will not be able to observe. Thus} \edit1{transit observations with} \jwst{} will be our best bet in the search for extrasolar life in the near term.

To build up sufficient signal as to be sensitive to atmospheric biosignatures, it will be necessary to stack dozens of transmission spectra from multiple observational epochs. Doing so will require sufficient understanding of the following concepts:
\begin{enumerate}
    \item The impact of stellar variability on the measured transmission spectrum from star spots and faculae;
    \item The impact of planetary variability on transmission spectra;
    \item Whether these varying conditions will invalidate the method of stacking transmission spectra to enhance the detection significance of low signal-to-noise spectroscopic features; and
    \item Whether biosignature gases in a planet's atmosphere are detectable through transmission spectroscopy.
\end{enumerate}

The search for life with the transit method hinges on our understanding of the habitable zone around a star. Because current technologies, which make use of the transit method, provide the best data for short period planets, M-dwarf stars with their small size and close-in habitable zones are the best targets to search for life until space-based direct imaging is realized. Some M dwarfs are more active and flare more frequently than their larger G-dwarf siblings \citep{Gunther2020}. The Sun has long been our testing ground for understanding how stellar activity in its various forms impact our observations of exoplanets, particularly for radial velocity observations \citep{Haywood2020}. Naturally, the Earth-Sun system presents the best opportunity for assessing the magnitude of these impacts and testing the feasibility of our techniques to search for life.

The fundamental goal of detecting signs of life on Earth using methods designed for exoplanets is crucial. If we are unable to account for both stellar and planet variability when the ground truth is known, or effectively stack dozens of transmission spectra to derive the presence of weak molecular features, then we should conduct the search for extrasolar life with other techniques. \edit1{Work by \citet{Zellem2017} demonstrates that this is less of a concern at infrared wavelengths, yet there remains a concern for highly active stars, such as WASP-19, and at optical wavelengths \citetext{Vatsal Panwar, in prep}. Additionally, improper correction for stellar surface heterogeneities on transmission spectra in the near-infrared can stymie our interpretation of planetary atmospheric properties \citep{Iyer2020, Rackham2018}.}

There are a number of barriers to detecting the spectra of Earth-like worlds. In this paper, \edit1{we show how observations of the Earth in transit across the Sun can be used as a unique and high-fidelity proxy, in fact the critical example case, for potentially habitable transiting exoplanets.} In \autoref{sec:bio} we break down the unanswered questions around detecting life as we know it using the transit technique with large collecting area, space-based observatories, such as \jwst{}, and suggest the use of small satellites to derive these answers. In \autoref{sec:stellar}, we simulate the transit light source (TLS) effect \edit1{\citep{Rackham2018}} on Earth-Sun transmission spectra. In \autoref{sec:planet}, we discuss the potential effects of planetary variations and simulate the effects of refraction in transmission observations. In \autoref{sec:stack} we discuss the caveats to stacking transmission spectra and what remains untested about the technique prior to the launch of \jwst{}. Finally, we summarize our conclusions in \autoref{sec:conc}.

\section{Transiting Exoplanet Geometry and Subsequent Biosignature Detection} \label{sec:bio}
A transmission spectrum is a measure of the planet's apparent change in size as a function of wavelength. Light from the host star passes through the planet's atmospheric annulus where it interacts with atoms and molecules. The annulus becomes opaque (and the planet appears larger) at wavelengths where these chemical species strongly absorb in the planetary atmosphere. Transmission spectroscopy data are thus sensitive to relative chemical abundances and the presence of cloud or haze particles within the atmosphere. A primary transit occurs when a planet passes in front of its host star, thus blocking part of the star's light as seen by the observer. Transmission spectra for the Earth have generally been achieved through indirect means, by observing solar occultations \citep[e.g.][]{Macdonald2019} or via a tertiary body reflecting the filtered starlight \edit1{\citep[e.g.][]{Palle2009, Vidal-Madjar2010, Ugolnikov2013, Arnold2014, Yan2015, Kawauchi2018, Youngblood2020}}. While these clever approaches have allowed us to derive Earth's transmission spectrum, they are only the first step towards understanding exoplanet transmission spectra in practice. For example, the Sun and the Earth's temporal variations remain unaccounted for. A satellite-based instrument actually observing an Earth-Sun transit in a geometry analogous to that anticipated for exoplanet transits is the only way to accomplish a true solar system laboratory test benchmark.

The transit geometry of the Earth-Sun system from within the Solar System has slightly different geometry than from \edit1{beyond}. When seen as a point source, the fractional dip in light from the star is determined by the planet-to-star area ratio; however, when the star and planet are spatially resolved, the transit depth is determined by the bodies' angular radii. Derivations of the equations for transit depth and molecular feature size (Equations \ref{eqn:depth} and \ref{eqn:signal}) are identical. However, when the star and planet are spatially resolved, the absolute radii are replaced by the angular radii.  Importantly, the absolute scale height is also replaced by the angular scale height.

\begin{eqnarray}\label{eqn:depth}
\textrm{Transit Depth} = \frac{\theta_\mathrm{p}^2}{\theta_*^2} = \left ( \frac{\tan^{-1}( \frac{R_{\mathrm{p}}}{d})}{\tan^{-1}(\frac{R_*}{a + d})} \right )^2 \nonumber \\
\approx \frac{R_{\mathrm{p}}^2}{R_*^2} \left ( \frac{a+d}{d} \right )^2 
\end{eqnarray}
\begin{eqnarray}\label{eqn:signal}
\textrm{Feature Size}  \approx \frac{2\theta_H\theta_\mathrm{p}}{\theta_*^2} = \frac{2 \tan^{-1}(\frac{H}{d}) \tan^{-1}(\frac{R_{\mathrm{p}}}{d})}{\tan^{-1}(\frac{R_*}{a+d})^2} \nonumber \\
\approx \frac{2 H R_{\mathrm{p}}}{R_*^2} \left ( \frac{a+d}{d} \right )^2
\end{eqnarray}

\noindent Here, $\theta_\mathrm{p}$ is the angular radius of the planet, $\theta_*$ is the angular radius of the star, and $\theta_H$ is the angular scale height of the planet as seen from a spacecraft within the Solar System. These angular radii are calculated in terms of the planet radius $R_{\mathrm{p}}$, stellar radius $R_*$, and scale height $H$ using the planet-spacecraft distance $d$ and the planet's semi-major axis $a$. The choice of units is arbitrary, but should be consistent across parameters. Equations \ref{eqn:depth} and \ref{eqn:signal} can be approximated, as shown, in the limit of small angles as long as $R_{\mathrm{p}} \ll d$, $R_* \ll a + d$, and $H \ll d$. In this form, it is easy to see how resolved solar system transits reduce to the familiar exoplanet-analog forms in the limit where the distance from the observer to the planet greatly exceeds the planet's semi-major axis, $d \gg a$. 

\begin{figure}
    \includegraphics[width=\linewidth]{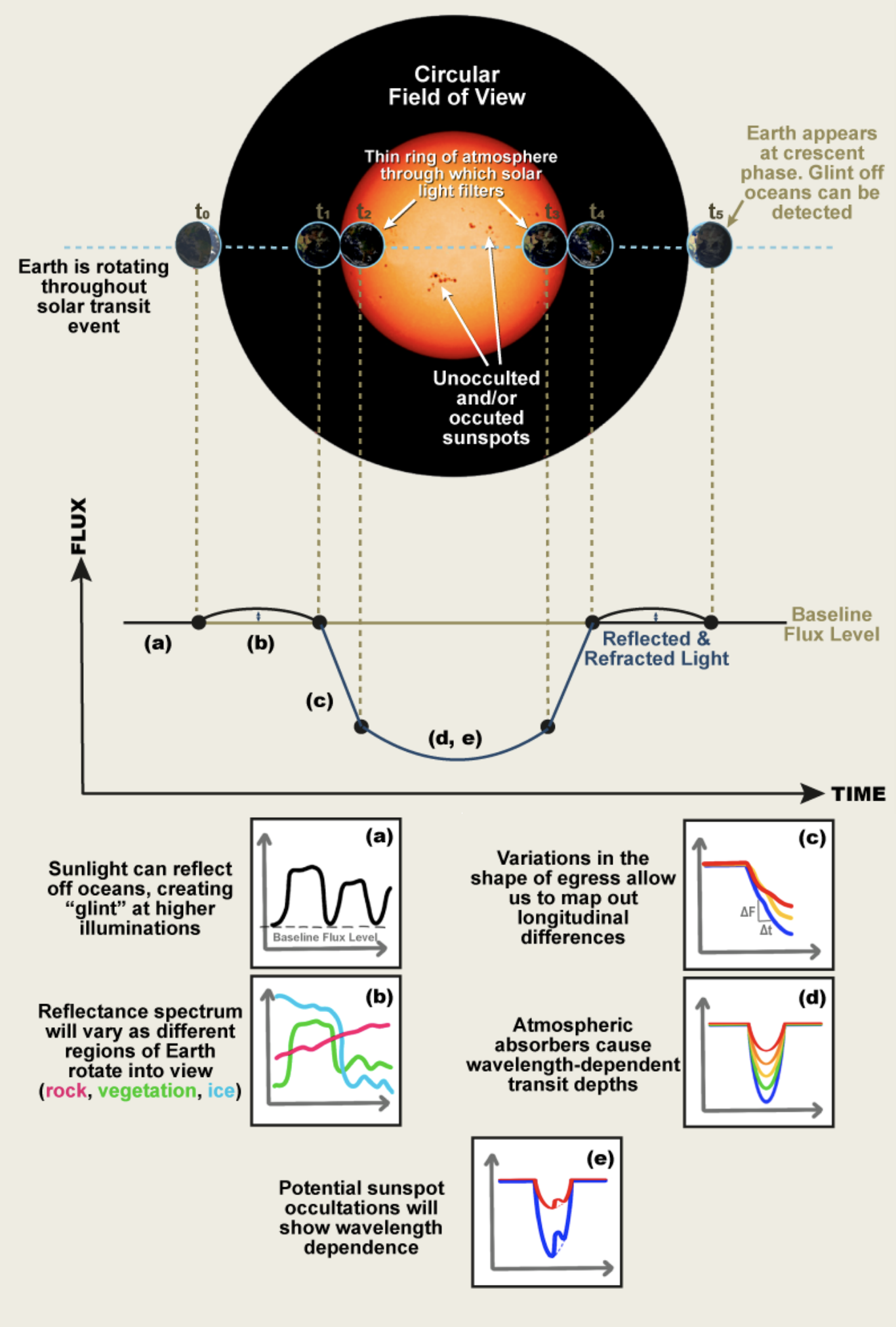}
    \caption{A schematic representation of a single Earth transit across the Sun as observed by a satellite beyond L2. We highlight several features that would be seen in these observations, including (a) ocean glint, outside of transit; (b) wavelength and temporal variability of reflected and refracted light; (c) ingress/egress as different longitudes and clouds rotate into view; (d) wavelength variability in transit depth due to absorption through and scattering by atmospheric constituents; and (e) stellar variability due to occulted and unocculted spots and faculae that can impact the measured transmission spectrum. For all sub-panels except (b), axes are flux vs. time and colors represent different wavelengths; for (b) the axes are flux vs. wavelength and the colors represent different contributions.}
    \label{fig:primer}
\end{figure}

A schematic diagram of the Earth transiting the Sun is shown in \autoref{fig:primer}. The total transit duration, \tdur{14}, begins at time \tdur{1} and ends at \tdur{4}, while the full transit duration, \tdur{23} spans \tdur{2}-\tdur{3}. The exact shape of the time-series observation is dictated by a variety of factors. Outside of transit, the baseline flux level is affected by the reflected light \deleted{ and refracted light} from Earth's atmosphere and surface \edit1{\citep[such as ocean glint, ][]{Robinson2010, Robinson2014b, Lustig-Yaeger2018}}, \edit1{and refracted light \citep{Betremieux2014, Betremieux2018, Misra2014}}. These signals are dependent on the precise orientation of Earth in its rotation and, thus, in theory longitudinal variations can be mapped by interpreting the variations in the shape of transit ingress and egress. In transit, each wavelength is affected by a different possible atmospheric absorber. The geometry as pictured in \autoref{fig:primer} requires the observer to be beyond the Earth-Sun L2 point, where \deleted{that} the apparent diameter of the Earth's disk is substantially smaller than that of the Sun\edit1{, the Sun and the Earth are the same angular diameter at 0.009\,AU, L2 is 0.01\,AU away, and the Moon is roughly 0.002\,AU away}. Spacecraft orbiting other Solar System planets may be able to achieve the requisite geometry, but these instruments risk damage from imaging an object as bright as the Sun. Also, not all spacecraft orbiting other Solar System bodies are able to observe Earth-Sun transits because the planets are not perfectly coplanar.

\begin{figure*}
    \centering
    \includegraphics[width=0.8\linewidth]{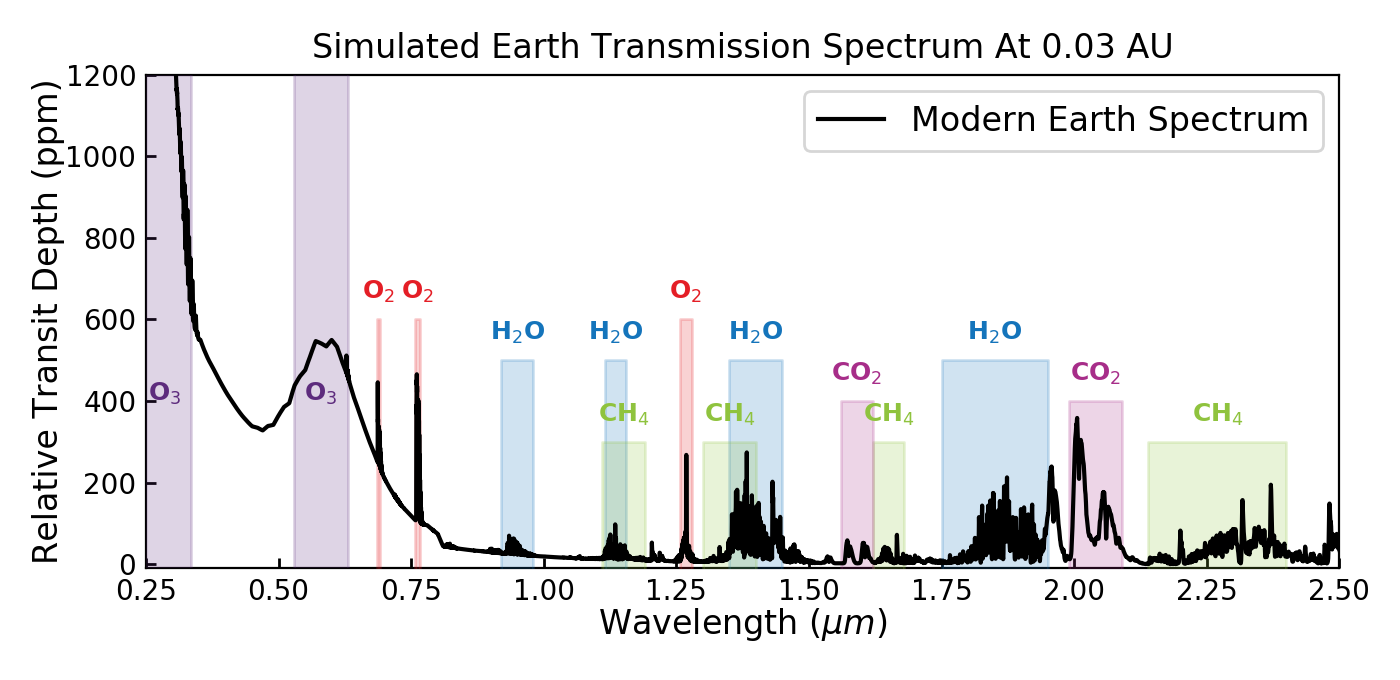}
    \caption{Simulated Earth transmission spectrum at a distance of 0.03\,AU from the Earth \citep[\edit2{originally generated by T.~Robinson}][]{Robinson2017, Robinson2018Handbook}.  Colored regions indicate wavelength ranges for the labeled molecular absorption features.  A spacecraft with broad wavelength coverage would be sensitive to standard habitability indicators (\hho, \coo) and biosignature pairs (\oo{} \& \chhhh, \ooo{} \& \chhhh) that \jwst{} will search for in M-dwarf planet atmospheres.}
    \label{fig:spec}
\end{figure*}

\autoref{fig:spec} displays a model transmission spectrum of the Earth at a distance of 0.03\,AU \deleted{(the Moon is at roughly 0.002\,AU from the Earth and L2 is roughly 0.01\,AU away)}. 
\edit1{The model assumes a present-day, Earth-like atmospheric composition \edit2{\citep{Robinson2017, Robinson2018Handbook}} and was originally generated by T.~Robinson for a TRAPPIST-1e-like system configuration, but subsequently rescaled to meet our needs.} Using currently-available detectors with modest noise floors, molecules could be clearly detected and their relative abundances readily constrained. The limiting molecular absorber in terms of detectability is methane at 2.3\,\micron. The narrowest molecular feature that needs to be resolved spectrally is \oo{} at 0.76\,\micron{} (760\,nm). \edit1{The features of \ooo{} at 0.25\,\micron{} and \chhhh{} at 2.3\,\micron, sets the minimum and maximum wavelength range for observations. Additionally, observing wavelengths beyond 2.5\,\micron{} would require active cryogenic cooling \citep{Beletic2008}, which can be difficult in a small spacecraft form factor.}

\begin{figure}
    \includegraphics[width=\linewidth]{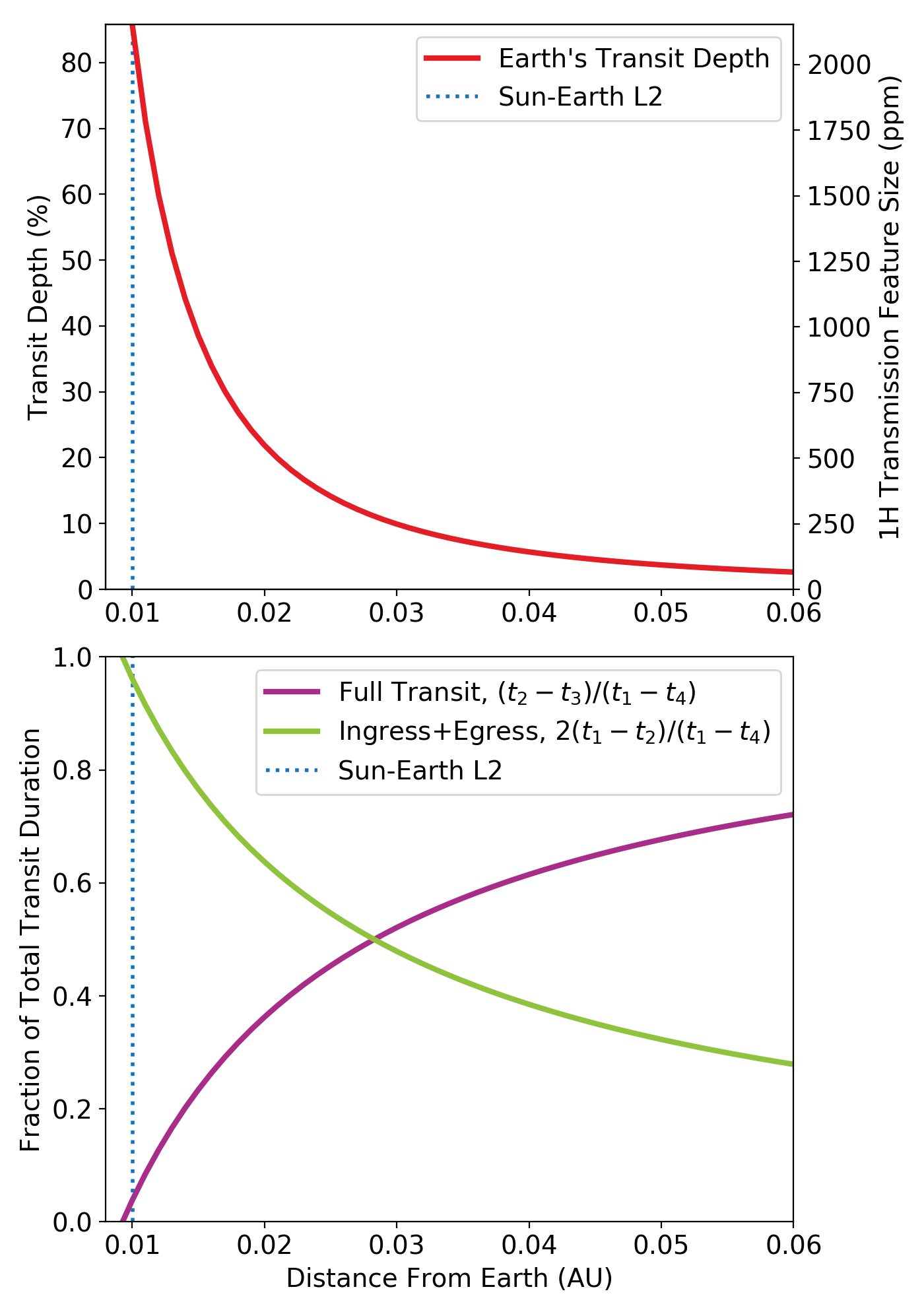}
    \caption{
    {\bf Top:} Earth-Sun transit depth and one-scale-height feature size versus distance from the Earth.  The reasonable transit depths and feature sizes just beyond the L2 Lagrange point imply that atmospheric characterization of a transiting Earth is achievable using currently available technology.  For example, at a distance of 0.03\,AU from the Earth, a satellite could observe transit depths of 10\% and measure spectroscopic features due to \hho, \oo, \ooo, \chhhh, and \coo{} with 1H signal sizes of 250\,ppm (see \autoref{fig:spec}).
    {\bf Bottom:} The fraction of total transit duration spent in full transit (violet) and ingress/egress (green) versus distance from the Earth.  A spacecraft closer than 0.02\,AU from the Earth may observe deeper transit depths with larger feature sizes; however, the full transit duration is significantly shorter and, thus, provides less information overall.  Distances of 0.02 -- 0.05\,AU from the Earth provide a good balance of measurable feature sizes and relatively long full transit durations.
    }
    \label{fig:scaleheight}
\end{figure}

The one-scale-height molecular feature size from the Earth's atmosphere as it transits the Sun would normally be very small (about 0.2\,ppm) if viewed from a neighboring stellar system; however, the close proximity of a satellite to the Earth presents a unique opportunity to boost the signal size by three orders of magnitude\edit1{, because the planet is proportionally larger in comparison to the host star}. The top panel of \autoref{fig:scaleheight} plots the computed transit depth and feature size as a function of distance from the Earth, while the bottom panel does the same for the fraction of total transit duration. Observing transits of the Earth-Sun system from closer than roughly 0.02\,AU is sub-optimal, because the full transit time is very short. At a distance of 0.03\,AU, the transit depth is 10\% and the 1H feature size is 250\,ppm. Although a 10\% transit depth may not seem analogous to a real \edit1{terrestrial} exoplanet observation, there is fundamentally no difference \edit1{from a Jupiter-sized planet observation} in this case once the planet is fully in transit (\tdur{2}-\tdur{3}). Furthermore, these 250\,ppm feature sizes are analogous to those seen in hot Jupiter atmospheres \edit1{\citep{Iyer2016}}. Therefore, the lessons learned from a sequence of satellite observations at distances greater than 0.02\,AU will apply directly to the thousands of known planet-hosting systems. \edit1{Such observations can be accomplished by a small, single instrument satellite capable of observing Earth transits across the Sun beyond the Earth-Sun L2 point in the 0.2-\edit2{\replaced{0.25}{2.5}} \micron{} range with a sufficiently sensitive spectrometer to capture key atmospheric constituents.}

\section{Stellar Variability} \label{sec:stellar}
The transit light source (TLS) effect \citep{Rackham2018,Rackham2019} can have a significant impact on the measured transmission spectrum due to photospheric heterogeneities (i.e., spots and faculae) either inside or outside the transit chord. If a star's spectrum inside the transit chord differs from the disk-integrated spectrum then that difference is imprinted on the planet's measured transmission spectrum. For cooler M dwarfs, this contamination can be up to ten times larger than the expected one-scale-height feature size for a rocky planet atmosphere and can lead to the false detection of chemical species such as water \citep{Rackham2018}.

\begin{figure}
    \includegraphics[width=\linewidth]{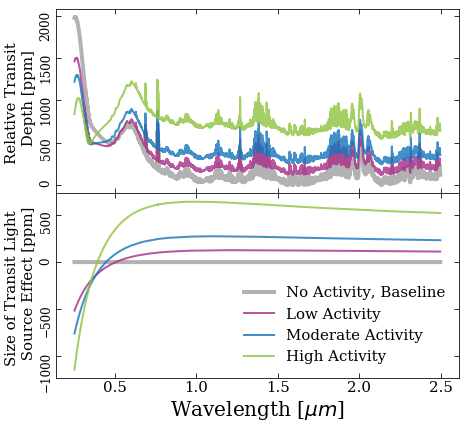}
    \caption{An example of the transit light source effect on an Earth-Sun transit observed by a satellite at a distance of 0.03\,AU. Top: The relative transit depth of the Earth depends on the amount of activity present on the Sun. Bottom: The size of the transit light source effect for varying levels of solar activity.
    }
    \label{fig:tls}
\end{figure}

\autoref{fig:tls} demonstrates the impact of the transit light source effect on a simulated observed transit. Following the formalism of \cite{Louden2017}, we assume the true transit depth is a function of the observed depth and the properties of the spotted regions of the stellar disk. We use a sample blackbody model for the solar, spot, and faculae component spectra. Typical spot and faculae levels for a Sun-like star are taken from \edit1{\citet{Rackham2019}}, which broadly agree with the activity levels seen on the Sun, and are used to calculate expected transit depth variations due to unocculted activity features during transit. \edit1{Following \citet{Rackham2019} we calculate the spot temperature as $T_{spot} = 0.418 \times T_{phot} + 1620 K$ and the faculae temperature as $T_{fac} = T_{phot} + 100 K$, assuming a stellar photosphere of 5800\,K. Our low, moderate, and high activity levels follow the bounds of 1$\sigma$ uncertainties on activity levels in \citet{Rackham2019} - specifically faculae covering fractions of {6\%, 10\%, 18\%}, respectively, with {0.6\%, 1.1\%, 2.2\%} covering fractions for the spot component.} The baseline transit spectrum is Earth as observed by a satellite at 0.03\,AU. While the TLS effect is an unconstrained problem in exoplanet observations, our knowledge of spot and faculae distributions on the Sun during observed transits will allow us to remove the TLS effect at high accuracy, while testing and comparing to methods used to account for stellar activity in exoplanet transits.

Observed variations in the measured exoplanet transit depth and shape are usually associated with unocculted and occulted stellar features, respectively \citep{Pont2007, Alonso2008, Knutson2011, Kreidberg2014, Bruno2018}. If not properly accounted for, stellar variability can impact derived parameters such as the orbital inclination, stellar density, and limb darkening coefficients. Numerous strategies exist to correct for stellar variability. For example, in their work deriving the transmission spectrum of TRAPPIST-1g, \citet{Wakeford2019} used out of-transit TRAPPIST-1 spectra to reconstruct the stellar flux using three model components ($T_\mathrm{eff}$ = 2400\,K, 35\% spot coverage at 3000\,K, and a 5800\,K hot spot covering \textless 3\% of the surface). An observation of an Earth-Sun transit has an advantage over these exoplanet observations, because Earth-based solar observatories are actively monitoring the Sun. To best test common methods of removing stellar variability in exoplanet transits, one can make use of the Sun's known spot and faculae distribution – regularly monitored and mapped by multiple existing Sun-observing spacecraft and instruments \edit1{such as space-based observations done with the Solar Dynamics Observatory \citep{Couvidat2016} and ground-based add-ons to HARPS \citep{Milbourne2019} and ESPRESSO \citetext{Nuno Santos, private communication}}– during each transit, as well as employ traditional exoplanet analysis methods involving photometric monitoring of the Sun outside of transit to derive activity levels.

\edit1{One such measure of activity is the presence of sunspots on the solar disk. }\autoref{fig:primer}, panel e, demonstrates an example of an occulted sunspot, where a ‘bump' in the transit light curve represents the temporary decrease in occulted light as the Earth passes over the dimmer region of the Sun \edit1{\citep{Miller-Ricci2008, Rabus2009}}. While occulted spots such as this make their presence known \edit1{and can be useful \citep{Sanchis-Ojeda2011, Sanchis-Ojeda2013}}, it is the unseen unocculted spots that require assumptions about spot coverage fractions and temperature differences for appropriate corrections in exoplanet observations. Because stellar activity can mimic atmospheric spectral features from the planet \edit1{\citep{McCullough2014, Sing2015, Bruno2020}}, it is crucial that their effects be accurately accounted for in exoplanet observations. An Earth-Sun transit data set facilitates the testing of the corrections applied to exoplanet transits and enables more effective strategies when faced with stellar variability from unresolved sources. For example, a first analysis of the data could be done with no knowledge of the spot locations, but, thanks to long-term studies of the Sun as a resolved star, the locations of solar activity from independent solar observations can be used to aid data analysis. Potential strategies might include \deleted{identifying more reliable spectral indicators of activity (between 0.25–2.5\,\micron),} improving occulted starspot modeling techniques and/or developing new strategies for transit spectroscopy observations of exoplanets using complementary instruments.

\section{Planetary Variability} \label{sec:planet}

Exoplanet transmission spectra are limb-averaged observations that contain information from an ensemble of light rays transmitted through all planetary latitudes \edit1{\citep{Feng2016, Feng2020, Pluriel2020}}, including contributions from the northern and southern hemispheres, and from clear and cloudy regions. These spatial differences evolve with time as the planet rotates and the seasons change. Thus, planetary variability poses a unique challenge for the accurate interpretation of heterogeneous Earth-like exoplanets that exhibit both spatial variabilities, blended within each transit observation, and temporal variabilities, from one observed transit to the next. A satellite at the proper distance to observe the full annulus of the Earth's atmosphere can directly probe the magnitude of planetary variability seen in the transmission spectrum of Earth and enable the development of retrieval methods for correctly interpreting exoplanet transit observations.

Earth's equator-to-pole temperature gradients and active hydrological cycle drive spatial variations that will be integrated together in the exoplanet-analog transmission spectra. Notably, Earth is a partially cloudy planet that permits both clear and cloudy optical paths through its atmosphere. These transit spectroscopy observations are necessary as they will be used to assess whether the combination of clear and cloudy optical paths allows for the accurate retrieval of the cloud-top pressure \edit1{\citep{Line2016a}}, and whether the inferred abundances of atmospheric gases such as \hho{} and \ooo{} are consistent with the range of spatial and vertical variation seen across the Earth.

Earth's axial tilt and geography creates measurable seasonal variability in the thermal structure, cloud patchiness, and relative abundance of trace atmospheric species. For example, Earth's integrated spectra are distinct at winter and summer solstice, due in part to the aggregation of land in its northern hemisphere \citep{Olson2018,Mettler2020}. While tidally-locked planets do not have seasons\edit1{\footnote{The timescale over which tidal obliquity erosion occurs predicts that axial tilts are unexpected for synchronously rotating planets \citep{Goldreich1966,Heller2011}.}}, it is important to quantify the magnitude of this effect in relation to other sources of planetary variability. 

\begin{figure}
    \centering
    \includegraphics[width=\linewidth]{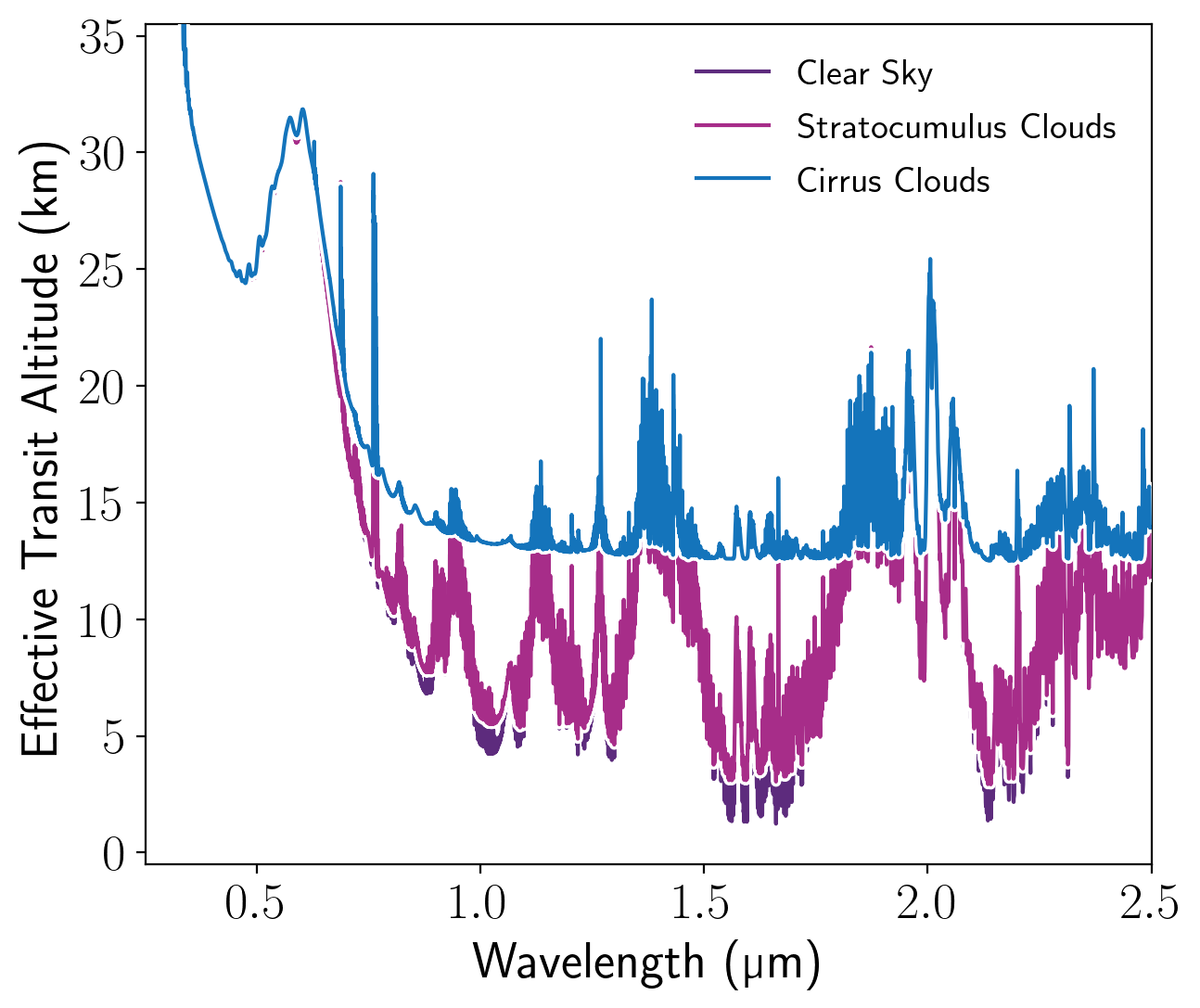}
    \caption{\edit1{Model transmission spectra of Earth for different end-member cloud cases. Earth without any clouds (purple) probes down to the near surface, while stratocumulus clouds (magenta) and cirrus clouds (blue) raise the observed continuum. An actual observation of Earth would be a combination of these end-members depending on Earth's orientation and particular weather patterns, such as seasonal developments of large storms.}}
    \label{fig:clouds}
\end{figure}

Earth is expected to exhibit some level of variability due to its rotation during a single transit \edit1{\citep{Palle2008}}. For example, a full transit lasting \edit1{six} hours in duration will sample half of the Earth's atmosphere. Thus, each \deleted{one-hour} segment can provide a semi-independent measurement of planetary variability as if they were subsequent transits form a tidally-locked M-dwarf planet. Furthermore, large variations in weather patterns, forest fires, volcanic activity, etc. can have a measurable impact on relatively short timescales. \edit1{In \autoref{fig:clouds} we show the transmission spectrum of the Earth with different atmospheric constituent end-members, i.e. as if the entire limb was 100\% clear, 100\% stratocumulus clouds, or 100\% cirrus clouds. \edit2{We model cirrus (ice) and stratocumulus (liquid) water clouds using the same Earth-validated model as \citet{Robinson2011}, which was based on spacecraft measurements, and is described in greater detail in \citet{Meadows2018} and \citep{Lincowski2018}. Cirrus clouds assume optical properties from B.~Baum's Cirrus Optical Property Library \citep{Baum2005}, and consist of a variety of particles including 45\% solid columns, 35\% plates, and 15\% 3D bullet rosettes, spanning 2--9500\,\micron{} with a cross section weighted mean diameter of 100\,\micron{}. Stratocumulus clouds assume Mie scattering optical properties with a two-parameter gamma distribution (a = 5.3, b = 1.1) and mean particle radius 4.07\,\micron{}, and use the refractive indices of water from \citet{Hale1973}. Based on approximate Earth averages \citep{Robinson2011}, cirrus clouds are placed near 8.5\,km altitude and have an optical depth of 3, while stratocumulus clouds are placed near 1.5\,km and have an optical depth of 10.} A satellite observation of Earth in transit would be a combination of these end-members depending on Earth's physical orientation and temporal variations in weather patterns, such as seasonal developments of large storms.} Thus, we find that it is necessary to observe multiple transits of Earth to measure the magnitude of these variations in Earth's transmission spectra and trace them back to specific sources via comparison with publicly available independent data acquired from Earth-observing satellites. These data will enable us to quantify the impact of planet variability on exoplanet transmission spectra and assess the feasibility of combining transit data over many years.

Recently, \citet{Macdonald2019} published an empirical clear sky transmission spectrum of Earth from 2-14\,\micron{} using solar occultation measurements. This was a first step to establishing such a data set to assess planetary variability. However, this study was limited by its 2--14\micron{} wavelength range, missing the UV and optical, and the omission of clouds to produce a clear sky transmission spectrum of Earth. The transmission spectrum of \citet{Macdonald2019} was unable to assess crucial regions of Earth's spectrum, because a much wider window is needed from the UV to the NIR to include strong \oo{} absorption (0.76 and 1.27\,\micron) and Rayleigh scattering. To augment the wavelength range observed previously and detect these important species, the UV and the optical must also be observed. Clouds are well known to play a crucial role in determining an exoplanet's transmission spectrum continuum, thereby limiting the depth into the atmosphere that can be probed \citep{Fortney2005,Kreidberg2014,Sing2016}. Limb-integrated observations of Earth as a true exoplanet transit analog are thus fundamentally necessary and will contain contributions from clear and cloudy optical paths. Because we can independently assess the state (e.g. cloudiness, etc.) of the Earth's atmosphere in a resolved sense, we can disentangle the contributions of different forms of terrestrial atmospheric variability from a disk-integrated spectrum.

\subsection{Refraction}
The effects of refraction were not directly observed in the \citet{Macdonald2019} transmission spectrum of Earth, because it was produced from solar occultation measurements from an Earth-orbiting satellite\edit1{, the Canadian low-Earth orbit satellite SCISAT}. As a result, it was able to probe deeper into the Earth's atmosphere (down to about 4\,km) than any distant exoplanet observer would be able to during transit due to the critical refraction limit \citep{Betremieux2014, Misra2014}. To mimic the effects of refraction on the spectral continuum analogous to terrestrial exoplanet observations around K and M dwarfs, an observer needs to be some distance from Earth. An exact Earth-Sun analog exoplanet system would have a refraction continuum floor at about 14\,km altitude above the solid surface, but for an Earth transiting an M dwarf, the refraction floor can drop below 10\,km \citep{Meadows2018, Lincowski2018}.

\begin{figure*}
    \includegraphics[width=\linewidth]{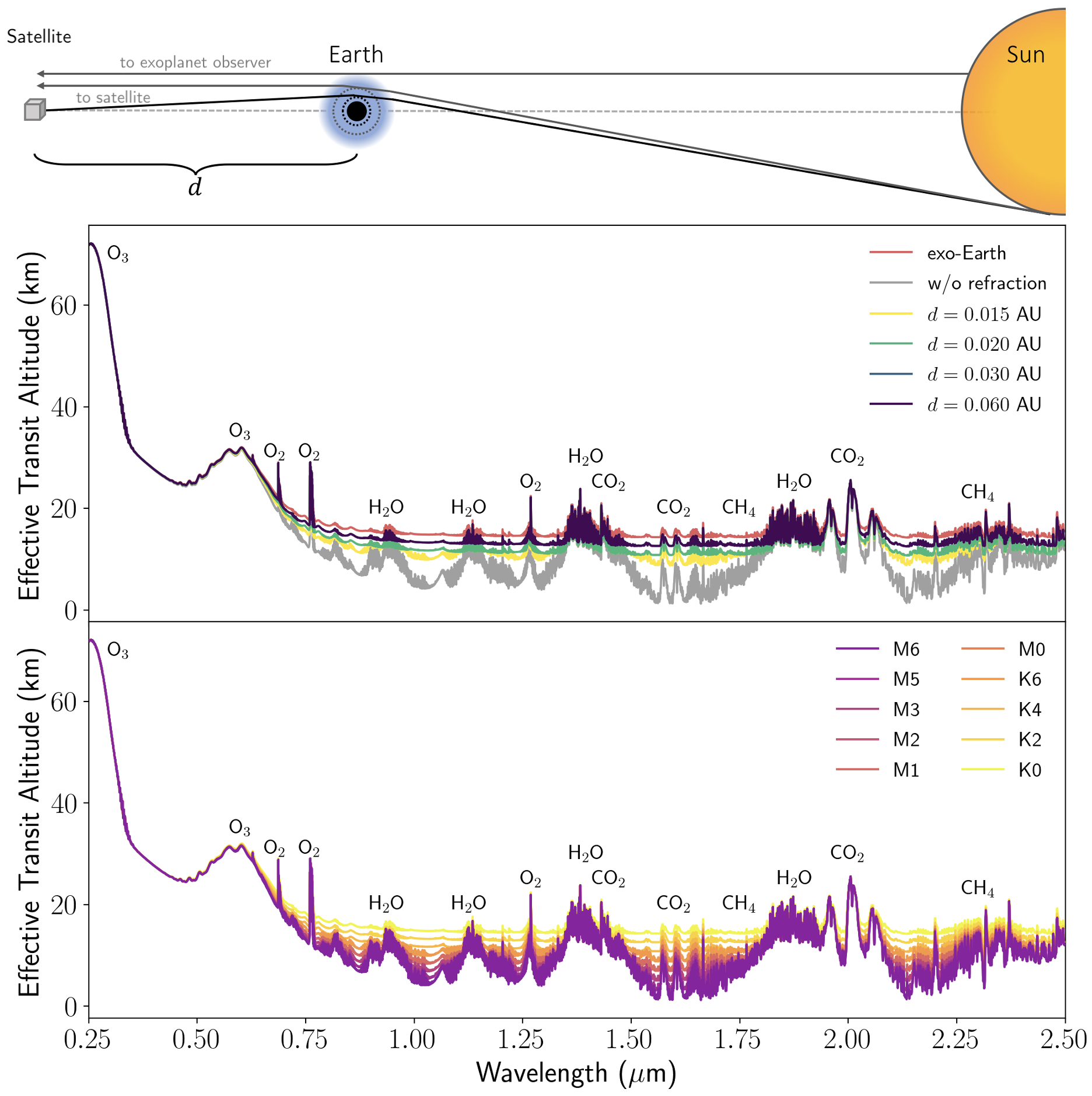}
    \caption{Refraction plays a crucial role in setting the transmission spectrum continuum for Earth and Earth-like exoplanets. Top: Solar rays from the limb of the Sun refract as they transmit through the Earth's atmosphere, creating a critical refraction limit (dotted circles) that restricts the depth probed into the atmosphere. The gray lines show an optical path and the refraction limit for an exo-Earth, while black lines show the same for a potential satellite at various Earth-spacecraft distances ($d$), which probes deeper due to the converging rays toward the spacecraft. Middle: Simulated transmission spectra with and without refraction for an exo-Earth (red and gray lines, respectively), which bracket the observed transmission spectra that vary with distance from Earth. Note that the line for 0.06\,AU is on top of the line for 0.03\,AU. Bottom: simulated transmission spectra of possible exo-Earths around various K and M dwarfs. Depending on the distance an Earth transit is observed from, the satellite can access regions of the Earth's atmosphere that are analogous to Earth-like planets transiting K and M dwarfs; such observations are highly relevant to future exoplanet observations. Note again that the M1 line is on top of M0 and M3 is on M2.}
    \label{fig:refract}
\end{figure*}

In \autoref{fig:refract}, we show the effect of refraction on an exo-Earth (an Earth-like planet at the inner edge of the habitable zone transiting various K and M dwarfs) observation as compared to an observation via satellite of Earth at various distances from 0.015\,AU to 0.06\,AU. Transmission spectrum models were produced using the Spectral Mapping Atmospheric Radiative Transfer (SMART) model \citep{Meadows1996} with the ray tracing refraction transmission spectrum upgrades described by \citet{Robinson2017}. We used a spatially averaged 1D atmospheric thermal structure and composition for Earth from \citet{Robinson2011} as an input into the spectral model, and we simulated exo-Earths orbiting various M dwarfs assuming each planet is located at the inner edge of the \citet{kopparapu_et_al2013} habitable zone for each respective star. Since the transmission spectrum code is intended to be used to simulate exoplanet observations at infinity, we modeled the convergence of deflected solar rays towards the satellite (as shown in the upper diagram in \autoref{fig:refract}) by simulating the atmospheric radiative transfer in a rotated reference frame. In the rotated reference frame, light rays may originate from beyond the solar disk and emerge from the Earth's atmosphere parallel as if heading to an observer at infinity. Thus, by artificially extending the solar radius by the exact amount necessary to preserve the maximum angle of deflection, we are able to simulate identical optical paths to a spacecraft observing the transit of Earth, assuming it lies along the Sun-Earth line to preserve symmetry.   

The advantage of a satellite observing a transit at a greater distance from Earth than 0.015\,AU would thus be to observe refraction continuum floors at slightly different altitudes as shown in \autoref{fig:refract}. These regions of the lower atmosphere are analogous to exoplanets transiting K- and M-dwarf stars. At 0.015\,AU away from the Earth, the refraction floor will be at an altitude of 8.7\,km, while at 0.06\,AU the floor will rise to 12.5\,km. These hard refraction boundaries may limit the ability to characterize the lower atmosphere of Earth; however, this is a necessary test to assess the capabilities of exoplanet transmission spectroscopy and inform future modeling work. Additionally, refraction is also expected to induce subtle time-dependent effects into Earth's transmission spectrum due to asymmetries in the atmospheric region probed, which changes depending on the exact alignment of the Earth and Sun \citep{Misra2014}. A satellite at \edit1{an appropriate} distance from the Earth \edit1{($\gtrsim$0.03\,AU)} enables a rigorous study into the numerous factors that inject variability into transmission spectrum observations.

\section{Stacking Transmission Spectra} \label{sec:stack}
\begin{figure}
    \includegraphics[width=\linewidth]{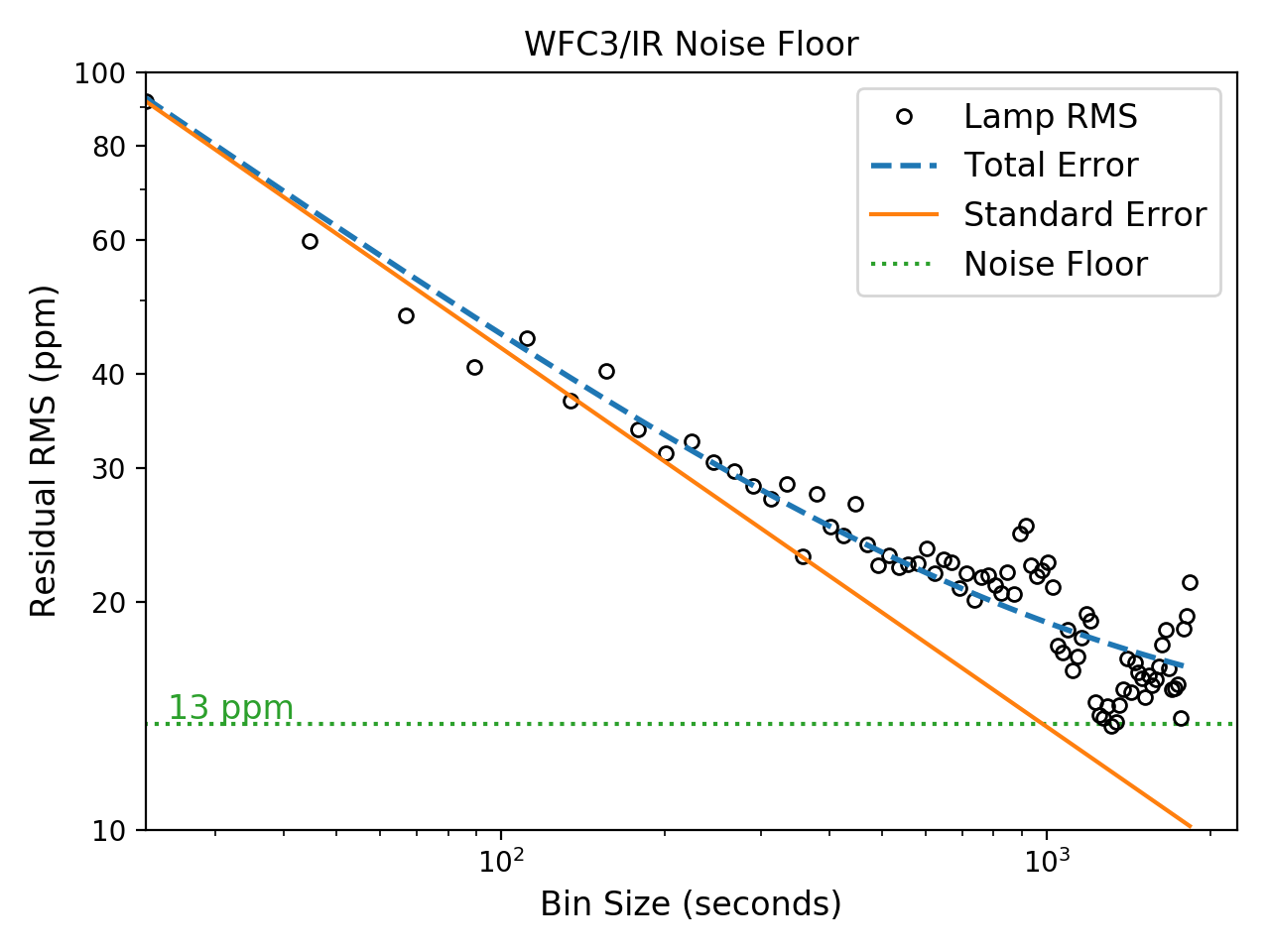}
    \caption{\edit1{Measured residual rms vs bin size (black circles) using time-series Tungsten lamp data obtained as part of {\hst} calibration program 15400.  At \edit2{\replaced{large}{small}} bin sizes, the measured rms follows the expected standard error (solid orange line).  At \edit2{\replaced{smaller}{larger}}  bin sizes, the rms begins deviating from the standard error.  The best-fit total error (dashed blue line) requires a noise floor of 13 ppm (dotted green line).}}
    \label{fig:noisefloor}
\end{figure}

With perfect detectors and no noise floor, stacking (or co-adding) exoplanet transmission spectra is a viable technique to improve precision and increase the detection significance of atmospheric constituents with weak spectroscopic features. The concept has previously been applied in limited capacities for transiting exoplanets. Notably, \citet{Kreidberg2014} obtained 15 transits of the sub-Neptune GJ~1214b using the WFC3 instrument on the \emph{Hubble Space Telescopes} \hst. However, the brightness of GJ~1214 did not push the limits of the detector down to its noise floor and the measured flat spectrum prohibited any constraints on the atmospheric composition. In reality, all detectors have inherent noise floors, and most are poorly understood. 
\edit1{For example, \cite{Greene2016} adopt noise floors of 20, 30, and 50 ppm for {\jwst}'s NIRISS, NIRCam, and MIRI instruments, respectively, whereas \citet{Schlawin2021} estimate a NIRCam noise floor of only 9 ppm.  Similarly for {\hst}, \citet{Stevenson2019a} analyzed eight years of WFC3 time-series observations to estimate a noise floor of $<21$ ppm at a resolving power of $\sim$40.  Using Tungsten lamp data from {\hst} calibration program 15400, we estimate a WFC3 noise floor of 13 ppm (see \autoref{fig:noisefloor}). While \citet{Zellem2017} modeled and analyzed stacking transits and the impact of epoch-to-epoch stellar variability, ultimately}, stacking transmission spectra from a well calibrated instrument to search for signals that are comparable in magnitude to a detector's noise floor is problematic and unproven.

Better characterization and enhanced detector stability is a long-recognized need. One of the primary challenges is in identifying and maintaining a sufficiently stable light source to conduct tests and calibrate instruments in the lab. The most promising lead is blackbody emission from a source that is temperature-controlled at the sub-mK level. The goal is to achieve \textless 10\,ppm stability over several hours (the typical duration of an exoplanet transit observation) and then co-add spectra from individual runs \edit1{\citep[e.g.][]{Tremblay2020}}. Therefore, the problem is a combination of not enough signal from a planet as well as too much noise that is poorly characterized.

The ideal solution is for a spacecraft to obtain many hours of transmission spectra from an extremely bright and photometrically-stable source, such as the Sun. In doing so, we could determine the limits of its detectors (just as exoplanet observations will test \jwst's detectors) and validate the process of stacking exoplanet transmission spectra to derive chemical abundances from constituents with spectral feature sizes comparable to the instrument noise floor. \citet{Morley2017} determined that 40 transits are sufficient to characterize planets such as GJ~1132b and TRAPPIST-1b. Observations of the Earth can achieve levels of precision that, to date, no exoplanet observation has achieved. Nor has a single planet's atmosphere yet been characterized using 40+ transit observations. Such exquisite transmission spectra of the Earth would validate the effectiveness of stacking transmission spectra and lead to establishment of best practices for such techniques for future missions. \edit1{GJ~1132b has a transit duration of $\sim$47 minutes \citep{Berta-Thompson2015}, and TRAPPIST-1b has a transit duration of $\sim$36 minutes \citep{Gillon2017}; therefore an assumption of a 1-hour transit is representative of this class of planet.}Thus, in order to simulate the stacking of 40 \deleted{1-hour} transits, a spacecraft would need to obtain 40+ hours of in-transit data \edit1{depending on the duration of each transit and how much of the Earth's rotation was captured}.

\section{Conclusions} \label{sec:conc}
\begin{figure*}[!h]
    \centering
    \includegraphics[width=\linewidth]{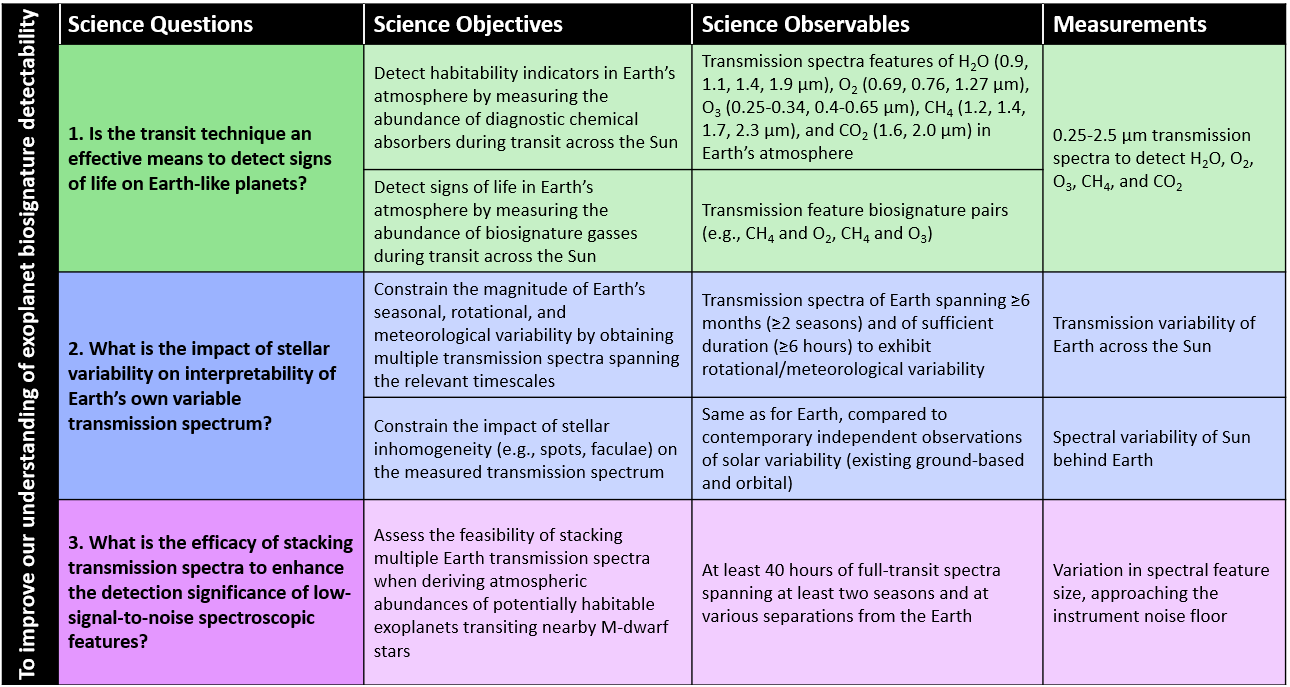}
    \caption{The high-level science questions pertaining to the detection of atmospheric biosignatures using the transit technique.  The corresponding science objectives can address these questions using a relatively small and inexpensive spacecraft by inducing Earth transits across the Sun.
    }
    \label{fig:stm}
\end{figure*}

In this paper, we have examined the prospects for solving three major challenges in detecting biosignature gases on terrestrial planet atmospheres: stellar variability, planet variability, and how these affect our fundamental technique of stacking spectra to build up the requisite signal-to-noise. We outline our resulting science questions and objectives in \autoref{fig:stm}.

We conclude that a true Earth-Sun transit observation is not only needed, but optimal for putting our techniques for exoplanet atmospheric characterization and biosignature detection to the test. An Earth-Sun transit data set \edit1{collected from a dedicated small satellite would} facilitate the testing and development of effective strategies to correct for stellar variability from unresolved sources. We also find that it is necessary to observe multiple transits of Earth to measure the magnitude of planetary variations in Earth's transmission spectra and trace them back to specific atmospheric and surficial sources. Limb-integrated observations of Earth as a true exoplanet transit analog are necessary to develop techniques to constrain spatial inhomogeneity in cloud cover. The transmission spectra of Earth by \citet{Macdonald2019} is able to probe deeper into the Earth's atmosphere than possible by an exoplanet observer, and thus these critical refraction boundaries have yet to be understood in the context of interpreting exoplanet spectra. We conclude that observing transits of the Earth-Sun system from further than L2 via a satellite capable of imaging the Sun are optimal for detection of biosignature gases on Earth at a similar geometry and signal size to that expected on extrasolar planets and the testing of the effectiveness of stacking transmission spectra.

\acknowledgements
The authors thank Justin Atchison (APL), and Jeff Rich and Darren Garber (Xplore). The majority of this work was funded by internal research and development funding from the Johns Hopkins Applied Physics Laboratory, including the Janney Energize Program. ECM is supported by an NSF Astronomy and Astrophysics Postdoctoral Fellowship under award AST-1801978. 

\software{Astropy \citep{astropy, astropy2},  IPython \citep{ipython}, LBLABC \citep{Meadows1996, Crisp1997}, Matplotlib \citep{matplotlib},  NumPy \citep{numpy, numpynew}, Pandas \citep{pandas2020, pandas2010}, SMART \citep{Meadows1996, Crisp1997}}

\bibliography{references}{}
\bibliographystyle{aasjournal}

\end{document}